\documentclass{book}
\usepackage[contrib,lang=american]{ems-book} 

\usepackage[capitalize,nameinlink]{cleveref}
\newtheorem{lemma}{Lemma}[section]
\newtheorem{conjecture}[lemma]{Conjecture}

\theoremstyle{definition}

\DeclareMathOperator*{\Tr}{{Tr}}
\newcommand{\rmdB}{\mathrm{d}B}
\newcommand{\rmdist}{\mathrm{dist}}
\newcommand{\cO}{\mathcal{O}}

\newcommand{\supp}{\mathrm{supp}}
\newcommand{\be}{\begin{equation}}
\newcommand{\ee}{\end{equation}}

\begin{document}
\mainmatter

\title{On Lieb-Robinson Bounds for the Double Bracket Flow}
\titlemark{On Lieb-Robinson Bounds for the Double Bracket Flow}

\emsauthor{1,2}{Matthew B.~Hastings}{M.~B.~Hastings}
                                                                                                                                                                                                                   \emsaffil{1}{Station Q, Microsoft Quantum, Santa Barbara, CA 93106-6105, USA}                                          \emsaffil{2}{Microsoft Quantum and Microsoft Research, Redmond, WA 98052, USA}


\keywords{Lieb-Robinson Bounds, Double bracket flow}

\begin{abstract}
We consider the possibility of developing a Lieb-Robinson bound for the double bracket flow\cite{brockett1991dynamical,chu1990projected}.  This is a differential equation
$$\partial_B H(B)=[[V,H(B)],H(B)]$$ which may be used to diagonalize Hamiltonians.  Here, $V$ is fixed and $H(0)=H$.
We argue (but do not prove) that $H(B)$ need not converge to a limit for nonzero real $B$ in the infinite volume limit, even assuming several conditions on $H(0)$. 
However, we prove Lieb-Robinson bounds for all $B$ for the double-bracket flow for free fermion systems, but the range increases \emph{exponentially} with the control parameter $B$.
\end{abstract}

\makecontribtitle

\section{Some Cases Where Lieb-Robinson Bounds Can and Cannot be Proven}
Lieb-Robinson bounds describe how an operator ``spreads out" over space while undergoing the Heisenberg equation of motion:
$$\partial_t O(t)=i[H,O(t)].$$
for some Hamiltonian $H$ on some quantum lattice system.
Crucially, one can bound this spread in a way that is uniform in the size of the system if $H$ obeys suitable locality properties, so that the spread depends only on the strength and range of terms in $H$.  Indeed, while the first proof\cite{lieb1972finite} also depended on the local Hilbert space dimension, later it was possible to derive bounds uniform in this dimension, as in \cite{hastings2004lieb}.

For local Hamiltonians, the spread is at most linear in time, governed by a Lieb-Robinson velocity.  However, for more general Hamiltonians with polynomially-decaying long-range interactions, one can derive in some cases a weaker result\cite{hastings2006spectral,foss2015nearly,tran2019locality,else2020improved} where the spread is superlinear in time, but one can still find bounds uniform in system size.

However, all these proofs use crucially the fact that the evolution considered is unitary.  If we instead consider an imaginary time version of the Heisenberg equation:
$$\partial_\tau O(\tau)=[H,O],$$
then it is not possible to derive bounds uniform in system size valid for long times.  
Indeed, a finite-time singularity may occur in the limit of infinite system size\cite{bouch2015complex}. 
The trouble is that for a system of volume $V$, an operator $O$ may have matrix elements between eigenstates whose energy differs by an amount of order $V$, and the imaginary time evolution causes these matrix elements to grow exponentially in $\tau$.
It is however possible to prove bounds for imaginary time evolution in
the case of one-dimensional systems\cite{araki1969gibbs,bouch2015complex}.
Also, for local Hamiltonians in arbitrary dimensions, one may use a series expansion\cite{bouch2015complex} to prove locality bounds which are valid for $O(1)$ time, where here $O(1)$ means a constant, independent of system size, but depending on microscopic parameters of the Hamiltonian.

Another case where bounds for imaginary time evolution can be proven is for free fermions.  Suppose each site $i$ has one Majorana operator $\gamma_i$, with
Majorana operators obeying canonical anticommutation relations, with
$$H=\sum_{i,j} \gamma_i h_{ij} \gamma_j,$$
where $h$ is some anti-symmetric anti-Hermitian matrix.
Suppose that there is some pseudometric $\rmdist(i,j)$ between sites, and suppose $h_{ij}=0$ for $\rmdist(i,j)>R$ for some range $R$ and $\Vert h \Vert \leq J$ where $\Vert \ldots \Vert$ denotes the operator norm. 
We have
$$\gamma_j(\tau)=\gamma_j + 2\tau \sum_k h_{kj} \gamma_k+\sum_{l,k}\frac{(2\tau)^2}{2!} h_{lk} h_{kj} \gamma_l+\ldots$$
The $m$-th term in this series is a linear combination of Majorana operators,
with the coefficients having $\ell_2$ norm bounded by
$(\tau J)^m/m!$ and hence the $m$-th term has operator norm bounded also by
$(\tau J)^m/m!$.  For $m\gg \tau J$, this is exponentially small in $m$, and hence up to exponentially small error, $\gamma_j(\tau)$ is supported within distance $O(R \tau J)$ of $j$.

\section{The Double Bracket Flow}
The double bracket flow\cite{brockett1991dynamical,chu1990projected} is a differential equation that can be used to diagonalize a Hamiltonian.  Let $V$ be some fixed Hermitian matrix and consider the equation
$$\partial_B H(B)=[[V,H(B)],H(B)],$$
with initial conditions $H(0)=H$, where $B\geq 0$ is a real parameter.   As $B\rightarrow \infty$, $H(B)$ converges to a fixed point, where it commutes with $V$.  Working in an eigenbasis of $V$ with eigenvalues of $V$ ordered in non-decreasing order so that $V=\mathrm{diag}(\lambda_1,\lambda_2,\ldots)$ with $\lambda_1\geq \lambda_2 \geq \ldots$, the only stable fixed points are where the eigenvalues of $H(B)$ are also ordered in non-decreasing order.

Defining $\eta(B)=[V,H(B)]$, then $-i\eta$ is Hermitian and the double bracket flow is $$\partial_B H(B)=i[\eta(B),H(B)].$$  So, it is equivalent to Heisenberg evolution of $H(B)$ under the ``Hamiltonian" $-i\eta(B)$ which itself depends on $H(B)$.

A closely related equation has been studied in physics\cite{wegner1994flow}, where it is sometimes called ``Wegner's flow equation".  The difference is that the Wegner's flow equation is defined by
$\partial_B H(B)=[[H_0(B),H(B)],H(B)],$ where $H_0(B)$ is the diagonal part of $H(B)$ in some basis.
This flow has also found applications in condensed matter physics\cite{kehrein2007flow}. 

Unfortunately, even if $V,H$ are both local, the flow generates increasingly complicated terms and in practice some truncation procedure is employed to 
approximate the evolution using a simpler local Hamiltonian.

So, it is of interest to see whether some kind of locality result analogous to a Lieb-Robinson bound might hold for this flow equation, to help control this truncation.  We discuss this in the next two sections.

Remark: the practical application of the double bracket flow is usually to study the \emph{ground state}.  In this case, in practice truncation procedures are employed which are (approximately) valid for the ground state.  We do not consider such issues at all here.

\section{Lieb-Robinson Bounds for the Double Bracket Flow: Local Spin Systems}
In this section, we consider arbitrary local spin systems.  We assume that there is some finite lattice of sites, $\Lambda$, with some metric $\rmdist(\cdot,\cdot)$, and the Hilbert space is a tensor product of finite-dimensional Hilbert spaces associated with each site.

Assume $H(0)=H$ and $V$ both are \emph{finite strength and range} with \emph{bounded local geometry}.  This means that $H=\sum_X h_X$ where each $X$ is supported on a set of diamter at most $R$ for some $R$ (this is the finite range), where $\Vert h_X \Vert \leq J$ for some $J$ (this is the finite strength) and where for any site $i\in \Lambda$, the number of sites $j$ with $\rmdist(i,j)\leq R$ is bounded by some constant (this is bounded local geometry).

These conditions are sufficient to prove a Lieb-Robinson bound for Heisenberg evolution under $H$.  We may then ask: does $H(B)$ obey some locality bound?  For any finite $\Lambda$, the evolution equation has a unique solution, but can one prove some bound on the locality of $H(B)$, uniformly in $|\Lambda|$?  

We attempt to prove such a bound in the next two subsections, but we are unable to do so, motivating the following conjecture (this conjecture clearly has several additional assumptions and does not in any way follow from our results but is motivated by them):
\begin{conjecture}
\label{conjbad}
There is some translation invariant Hamiltonian $H$ on a $d$-dimensional hypercubic lattice, and some other translationally invariant $V$, so that, if one considers these $H$ and $V$ on a sequence
of lattices of increasing size, the solution of the double bracket flow does not converge to a translationally invariant Hamiltonian in the infinite system size limit for any real $B \neq 0$.

Indeed, the solution may fail to converge even if the Hilbert space on each site is two-dimensional, $V$ is the sum of Pauli $Z$ on each site, and $H=V+\epsilon \Delta$ where $\Delta$ is some translationally invariant Hamiltonian and $\epsilon$ is an arbitrarily small real scalar.
\end{conjecture}

Remark:
As is known, the double bracket flow is a gradient flow.  Consider the function $\Tr((H(B)-V)^2)$.  Let $H(B)=U(B) H(0) U(B)^\dagger$ for unitary $U(B)$.  
Consider infinitesimal change in $U(B)$: $U(B)\rightarrow U(B)+\eta U(B)$ for infinitesimal anti-Hermitian $\eta$.  The change in the potential is then $-2\Tr([\eta,H(B)]V)=-2\Tr(\eta [H(B),V])$.
So, for the gradient flow on the manifold of $U(B)$ using a metric\footnote{Here we mean a metric on $U(B)$ rather than on $H(B)$.} induced by the Hilbert-Schmidt inner product is the double bracket flow.
So, an interesting open question is whether some modification of the double bracket flow can be defined which will have some locality properties.  It seems that one could replace the Hilbert-Schmidt inner product by some other inner product to improve the locality.  For example, on a system of qudits, an orthonormal basis for the Hilbert-Schmidt inner product may be obtained by products of nontrivial (generalized) Pauli operators on sets of sites $X$.  We might change the metric so that it is still diagonal in this basis, but scales with the diameter of $X$.

\subsection{Power Series}
\label{ps}
One might first try to prove a locality bound for $B=O(1)$ by the same 
kind of power series expansion technique that works for imaginary time Hamiltonian evolution\cite{bouch2015complex}.  Unfortunately, this does not seem to work for the double bracket flow as the expansion seems to have vanishing radius of convergence in the infinite system size limit.  Let us see what happens.  

The calculation here is not intended to be a proof; rather, we show that some subset of terms in a power series expansion may diverge for all $B\neq 0$, at least under some worst case assumptions on certain commutators.
It is possible that a more refined power series analysis might be able to prove a nontrivial result though; for example, perhaps these terms are canceled by other terms or the worst case commutation bound may not hold.  Thus we leave as an interesting open question: does the power series have a nonzero radius of convergence for some lattices in the infinite size limit, such as one-dimensional systems?

Let $H$ and $V$ both be a sum of local terms on some quantum lattice system.
We can expand $H(B)$ as a power series in $B$ for a finite lattice system.
Let $H_k(B)$ denote the $k$-th order term in $B$ for $H(B)$ so that $H_0(B)=H$.
We then have
$$\partial_B H_{k+1}(B)=\sum_{l=0}^k [[H_{k-l}(B),V],H_{l}(B)].$$

Let us consider a particular term in the sum, where $l=0$.  So we have
\be
\label{quoteq}
H_{k+1}(B)=\int_0^B [[H_k(B),V],H_0(B)]\rmdB +\mathrm{other \, terms}.
\ee

Suppose each term in $H_0$ and each term in $V$ is supported on some set of diameter $O(1)$ and has norm at most $J$.  Then, $H_k$ is supported on some set of diameter $O(k)$ and for some lattices, the double commutator may be of order $J^2 k^2$.
Let $J_k(B)$ be a bound on the norm of these terms of diameter $O(k)$ in $H_k$.
Precisely, we imagine that $H_k$ is a sum over sites $j$ of some term with norm at most $J_k(B)$ supported within diameter $O(k)$ of $j$.

Then we have
$$J_{k+1}(B)\leq \int_0^B J_{k}(B') J^2 k^2 |\rmdB'|,$$
ignoring the contribution of these other terms.
Replacing the inequality $\leq$ with an equality sign, and taking $J_0(0)=1$, $J_k(0)=0$ for $k>0$, the equation is easy to solve with
$$J_k(B)=\frac{(J^2 |B|)^k}{k!} ((k-1)!)^2,$$
and so $\sum_k J_k(B)$ diverges at any $B\neq 0$.

\subsection{Weak Perturbations}
\label{lrdbf}
Having failed to prove convergence of the power series for any $B\neq 0$, let us try something weaker.  Throughout this subsection we consider real, positive $B$.

Suppose $H$ and $V$ both have finite strength and range.  Further, suppose that
$H(B)=V+\epsilon \Delta(B)$ for some small nonzero $\epsilon$ and some $\Delta(B)$ where $\Delta(0)$ which has finite strength and range.
Then, the double bracket flow becomes
\be
\label{perturbdbf}
\partial_B \Delta(B)=[[V,\Delta(B)],V]+\epsilon [[V,\Delta(B)],\Delta(B)].
\ee
The first term is a linear term, with all eigenvalues nonpositive, while the second is nonlinear.

If we consider some simple case where, for example, we have a lattice of qubits and $V$ is a sum of Pauli $Z$ operators on each site, then the linear equation of motion can be solved readily.  Let $a_j=(X_j+iY_j)/2$ where $X_j,Y_j$ are Pauli $X,Y$ operators on site $j$.
We can expand $\Delta(B)$ of $B$ in a basis of products of single site operators $Z, a, a^\dagger$.  Given such a product in which $a$ appears a total of $n_+$ times and $a^\dagger$ appears a total of $n_-$ times, say that term has ``charge $q$" where $q=n_+-n_-$.
Then, any term with charge $q$ is an eigenoperator of \cref{perturbdbf} with
eigenvalue $-4 q^2$.

So, one might hope in this example that terms with with large $q$ will decay rapidly as $B$ increases, while terms with small $q$ operators will have a small commutator with $V$.  So, one might hope some $H(B)$ will obey some locality properties, uniform in $|\Lambda|$.

Unfortunately, this does not seem to hold.  Similar to the previous section, the argument here involves focusing on certain terms in a power series expansion.  It is not intended to be a proof, but it is intended to give some evidence to support \cref{conjbad}.

We solve \cref{perturbdbf} as a power series in $\epsilon$.
Let $\Delta_k(B)$ denote the term of order $\epsilon^k$ for $\Delta(B)$.
We then have
$$\partial_B \Delta_{k+1}(B)=[[V,\Delta_{k+1}(B)],V]+
\epsilon \sum_{l=0}^k [[\Delta_{k-l}(B),V],\Delta_{l}(B)].$$
Further, let $\Delta_{k,q}$ denote the terms with given charge $q$ in $\Delta_k$.
Then
$$\partial_B \Delta_{k+1,q}(B)=-4q^2\Delta_{k+1,q}(B)+
\epsilon \sum_{l=0}^k \sum_r[[\Delta_{k-l,q-r}(B),V],\Delta_{l,r}(B)].$$
Let us focus on terms with $l=0$ and $r=0$, writing
\be
\label{linperteq}
\partial_B \Delta_{k+1,q}(B)=-4q^2\Delta_{k+1,q}(B)+
\epsilon [[\Delta_{k,q}(B),V],\Delta_{0,0}(B)]+\mathrm{other \, terms}.
\ee

The term $\Delta_{0,0}(B)$ is independent of $B$.
Ignoring the other terms in \cref{linperteq}, this gives a linear equation
of motion for $\Delta_{k+1,q}$ in terms of $\Delta_{k+1,q}$ and $\Delta_{k,q}$.
The term $\Delta_{k,q}(B)$
is supported on some set of diameter $O(k)$ and for some lattices, the double commutator may be of order $\epsilon kqJ^2$, where $J$ is proportional to the strength of $\Delta(0)$.
Let $\delta_{k,q}(B)$ be a bound on the norm of these terms in $\Delta_{k,q}$.
Thus, in worst case we might expect a linear equation describing this norm:
$$
\partial_{B} \delta_{k+1,q}(B)=-4q^2 \delta_{k+1,q}(B) + \epsilon k qJ^2 \delta_{k,q}(B).$$
Equivalently, defining $$\tilde \delta_{k+1,q}(B)=\exp(4q^2B ) \delta_{k+1,q}(B),$$
we get
$$
\partial_{B} \tilde \delta_{k+1,q}(B)=\epsilon k q J^2\tilde \delta_{k,q}(B).$$
This equation is readily solved with
$\tilde \delta_{k,q}(B)=\frac{(k-1)!}{k!} (\epsilon qJ^2)^k B^k,$
and so the sum over $k$ has a finite-$B$ singularity at $B=1/\epsilon q J^2$.
Hence, $\delta_{k,q}$ \emph{also} has a finite-$B$ singularity at
$$B_q\equiv (\epsilon q J^2)^{-1},$$ at least in the approximations of this subsection.

We may take $q$ large to get a finite-$B$ singularity at arbitrarily small $B$.  Of course, the reader may object here: assuming $\Delta$ has finite range, then
$\Delta_{0,q}$ is nonvanishing only for charge $q=O(1)$.  However, starting with terms
with charge which is $O(1)$, then \cref{perturbdbf} will generate terms with any given $q$ for any nonzero $B$, with a strength proportional to $(\epsilon J^2B)^q/q!$. So, indeed we still expect a finite-$B$ singularity at the given $B_q$, and we may take $q$ large to get a singularity at arbitrarily small $B$.

\section{Free Fermions}
\label{ff}
One case where we can prove a kind of Lieb-Robinson bound for all $B$ is when $H(B)$ and $V$ both describe free fermions, with
$$H(B)=\sum_{i,j} \gamma_i h(B)_{ij} \gamma_j,$$
where $h(0)=h$,
and
$$V=\sum_{i,j} \gamma_i v_{ij} \gamma_j,$$
for some anti-symmetric anti-Hermitian matrices $h(B),v$.

Then,
$$\partial_B h(B)=4[[v,h(B)],h(B)].$$
Suppose that $h_{ij}=0$ for ${\rm dist}(i,j)>R$ and $\Vert h \Vert = J$ for some $R,J$ and similarly
$v_{ij}=R$ for ${\rm dist}(i,j)>0$ and $\Vert v \Vert = J$.

Define for any matrix $m$ the quantity
$\Vert m \Vert_r$ by
$$\Vert m \Vert_r \equiv {\rm max}_{\psi,\phi, {\rm dist}(\supp(\psi),\supp(\phi))>r} \langle \psi,m \phi\rangle,$$
where $\psi,\phi$ are vectors with $\ell_2$ norm $1$ and $\supp(\cdot)$ denotes the support of a vector.
Note that $\Vert \cdot \Vert_r$ is not a norm: while it obeys a triangle inequality, $\Vert m \Vert_r$ may be vanishing even if $m$ is nonvanishing.
Rather, $\Vert m \Vert_r$ may be thought of as measuring the strength of $m$ at distance $r$.

Our main result is \emph{there is a light-cone, but the excitations may spread exponentially quickly in $B$}.
In particular
\begin{lemma}
\label{expspread}
For any $\ell>0$, we have
$$\Vert h(B) \Vert_\ell \leq J 
\Bigl(\frac{8e J^2 B}{\log_2(\ell)-O(1)}\Bigr)^{\log_2(\ell)-O(1)}.$$

So, for $\log_2(\ell)$ large compared to $8e J^2 B$, the quantity
$\Vert h(B) \Vert_\ell$ is exponentially small in $\log_2(\ell)$.
\begin{proof}
For any $\psi,m,r$, we have
$m\psi=\psi_1+\psi_2$ where $\psi_1$ is supported within distance $r$ of
$\supp(\psi)$ with $|\psi_1| \leq \Vert m \Vert \cdot |\psi| $ and
where $|\psi_2|\leq \Vert m \Vert_r \cdot |\psi|$.
So, we have in general
\be
\label{distbound}
\Vert m_1 m_2 \Vert_{r_1+r_2} \leq \Vert m_1 \Vert_{r_1}\cdot \Vert m_2 \Vert + \Vert m_1 \Vert \cdot \Vert m_2 \Vert_{r_2}.
\ee

We have $\Vert h(B) \Vert=\Vert h \Vert = J$ for all $B$ since the flow equation describes a unitary evolution.
Hence, $\Vert h(B) \Vert_r \leq J$ for all $B,r$.

We have, for small $\rmdB>0$:
\be
\langle \psi,h(B+\rmdB) \phi\rangle= \langle \psi,h(B) \phi\rangle
+4 \langle  \psi,[[v,h(B)],h(B)] \phi \rangle \rmdB
+ \cO(\rmdB^2).
\ee
Let $|\psi|=|\phi|=1$.
Suppose $\rmdist(\supp(\psi),\supp(\phi))=2\ell+R$ for some $\ell$.
The term
$\langle  \psi,[[v,h(B)],h(B)] \phi \rangle$ is a sum of four terms, given by the different ways of expanding the commutator.  Consider a typical term such as
$\langle \psi, v h(B) h(B) \phi \rangle$.
Applying \cref{distbound} and using
$\Vert v \Vert_R=0$, we have $$\Vert v h(B) \Vert_{\ell+R}\leq J \Vert h(B) \Vert_\ell$$ and
\begin{align}\Vert v h(B) h(B) \Vert_{2\ell+R}&\leq \Vert v h(B) \Vert_{\ell+R} \cdot \Vert h(B)\Vert+\Vert v h(B) \Vert \cdot \Vert h(B) \Vert_\ell \\
\nonumber
&\leq 2J^2 \Vert h(B) \Vert_\ell.
\end{align}
Summing over the four terms, we find
\be
|\langle  \psi,[[v,h(B)],h(B)] \phi \rangle| \leq 8J^2 \Vert h(B) \Vert_\ell,
\ee
so
\be
\Vert h(B+\rmdB) \Vert_{2\ell+R} \leq \Vert h(B) \Vert_{2\ell+R} +8J^2 \Vert h(B) \Vert_\ell \rmdB+O(\rmdB^2).
\ee

Define a sequence of scales $R_k$ for $k=0,1,2,\ldots$ by
\begin{align}
R_0&=R \\ \nonumber
R_{k+1}=2R_k+R,
\end{align}
and taking a limit as $\rmdB\rightarrow 0^+$,
we have
\be
\label{recurrence}
\Vert h(B) \Vert_{R_{k+1}} \leq \int_0^B 8 J^2 \Vert h(B) \Vert_{R(k)},
\ee
with
also
$$\Vert h(B) \Vert_{R_k}\leq J$$
for all $k$.

Replacing the inequality in \cref{recurrence} with an equality, the result is easily solved since it is linear in the quantities $\Vert h(B) \Vert_{R(i)}$
and we find
\begin{align}
\label{stirling}
\Vert h(B) \Vert_{R_{k}} &\leq  J \frac{(8J^2 B)^k}{k!} \\ \nonumber
&\leq J \Bigl(\frac{8e J^2 B}{k}\Bigr)^k,
\end{align}
where we used Stirling's approximation.
Hence, for $k\gg 8eJ^2B$ for some $c>0$, it is exponentially small.

We claim that
$$R_k \leq (2^{k+1}-1) R.$$  This follows inductively from
$R_0\leq (2^1-1)\cdot R$
and $R_{k+1}=2R_k+R\leq (2^{k+2}-2)R+R=(2^{k+2}-1)R.$
Hence, given any $\ell$, we may choose
$k=\log_2(\ell)-O(1)$ with $R_k\leq \ell$.

So, from \cref{stirling}, 
$$\Vert h(B) \Vert_\ell \leq J 
\Bigl(\frac{8e J^2 B}{\log_2(\ell)-O(1)}\Bigr)^{\log_2(\ell)-O(1)}.$$
\end{proof}
\end{lemma}

We expect that the bounds of this lemma can be tightened significantly for $\ell$ large compared to this ``exponential light-cone".
However, the exponential spreading of \cref{expspread} is the best possible in general.  As a model system, consider the family of Hamiltonians $h_t$
in one-dimension, depending on a parameter $t$, with sites indexed by an integer $k$, with matrix elements
\begin{align}
(h_t)_{2k,2k+1}&=(h_t)_{2k+1,2k}=1+t, \\ \nonumber
(h_t)_{2k+1,2k+2}&=(h_t)_{2k+2,2k+1}=1-t,
\end{align}
with all other matrix elements vanishing.

Introduce a basis of Fourier modes $\sum_k \exp(i\theta) |k\rangle$ for $\theta$ periodic mod $2\pi$.
Then, in this basis, $h_t$ is a block two-by-two matrix for each pair $\theta,\theta+\pi$:
$$\begin{pmatrix}
2\cos(\theta) & t \\
t & -2\cos(\theta)\end{pmatrix}=t \sigma_x + 2 \cos(\theta) \sigma_z,$$
where $\sigma_x,\sigma_z$ are Pauli sigma matrices.

Let us pick some given $t$, and let $h=h_t$ and $v=h_{-t}$.
Each such two-by-two matrix can be evolved independently under the double bracket flow.
Writing $h_t(B)=a(B) \sigma_x + b(B)\sigma_z$ in a given block for given $\theta$,
the differential equation in a two-by-two block for $a(B),b(B)$ may be computed explicitly.  This evolution has two fixed points: when $h_t(B)=v$ or when $h_t=-v$.  The first fixed point is stable while the second is unstable.  Any perturbation of the unstable fixed point grows exponentially in $B$.

However, for $\theta$ near $\pm \pi/2$, we have $\cos(\theta) \approx 0$, and so we are near the unstable fixed point.  Indeed, at large $B$, for $|\log(\cos(\theta)|)$ large compared to $1/B$, then $h_t(B)$ evolves to $v$ in the given block, while for $\log(|\cos(\theta)|)$ small compared to $1/B$, then
$h_t(B)$ is close to $-v$.

So, $h_t(B)$ changes very rapidly as a function of $\theta$ for $\log(|\cos(\theta)|)$ of order $1/B$.  This corresponds to an exponential growth in $B$ of the appropriate length scale in $h_t(B)$ after Fourier transforming back to the lattice.

\bibliographystyle{emss}
\bibliography{refs}
\end{document}